\documentclass[12pt]{article}
\usepackage{latexsym}
\usepackage{amsmath}
\usepackage{amssymb}
\catcode`\@=11
\newcount\hour
\newcount\minute
\newtoks\amorpm \hour=\time\divide\hour by 60\minute
=\time{\multiply\hour by 60 \global\advance\minute by-\hour}
\edef\standardtime{{\ifnum\hour<12 \global\amorpm={am}%
        \else\global\amorpm={pm}\advance\hour by-12 \fi
        \ifnum\hour=0 \hour=12 \fi
        \number\hour:\ifnum\minute<10
        0\fi\number\minute\the\amorpm}}
\edef\militarytime{\number\hour:\ifnum\minute<10
0\fi\number\minute}
\def\draftlabel#1{{\@bsphack\if@filesw {\let\thepage\relax
   \xdef\@gtempa{\write\@auxout{\string
      \newlabel{#1}{{\@currentlabel}{\thepage}}}}}\@gtempa
   \if@nobreak \ifvmode\nobreak\fi\fi\fi\@esphack}
        \gdef\@eqnlabel{#1}}
\def\@eqnlabel{}
\def\@vacuum{}
\def\marginnote#1{}
\def\draftmarginnote#1{\marginpar{\raggedright\scriptsize\tt#1}}
\def\draft{
        \pagestyle{plain}
        \overfullrule=2pt
        \oddsidemargin -.1truein
        \def\@oddhead{\sl \phantom{\today\quad\militarytime} \hfil
        \smash{\Large\sl DRAFT} \hfil \today\quad\militarytime}
        \let\@evenhead\@oddhead
        \let\label=\draftlabel
        \let\marginnote=\draftmarginnote
        \def\ps@empty{\let\@mkboth\@gobbletwo
        \def\@oddfoot{\hfil \smash{\Large\sl DRAFT} \hfil}
        \let\@evenfoot\@oddhead}
        \def\@eqnnum{(\theequation)\rlap{\kern\marginparsep\tt\@eqnlabel}%
        \global\let\@eqnlabel\@vacuum}  }

\renewcommand{\theequation}{\thesection.\arabic{equation}}
\renewcommand{\thefootnote}{\fnsymbol{footnote}}
\newcommand{\newsection}{    
\setcounter{equation}{0}\section}
\def\appendix#1{\addtocounter{section}{1}\setcounter{equation}{0}
\renewcommand{\thesection}{\Alph{section}}
\section*{Appendix \thesection\protect\indent \parbox[t]{11.15cm}{#1}}
\addcontentsline{toc}{section}{Appendix \thesection\ \ \ #1}}

\def \bi{\bibitem}
\def \la {\label}

\def \b {\beta}

\def\be{\begin{equation}}
\def\ee{\end{equation}}

\catcode`\@=11

\def\bea{\begin{eqnarray}}
\def\eea{\end{eqnarray}}
\def\beann{\begin{eqnarray*}}
\def\eeann{\end{eqnarray*}}
\def\beq{\begin{equation}}
\def\eeq{\end{equation}}
\def\ba{\begin{array}}
\def\ea{\end{array}}
\def\ben{\begin{enumerate}}
\def\een{\end{enumerate}}

 \def \la {\label}
 \def\be{\begin{equation}}
\def\ee{\end{equation}}

\def \la {\label}

\font\mybb=msbm10 at 11pt

\def\bb#1{\hbox{\mybb#1}}

\def\bR {\bb{R}}

\def \ee {\epsilon}

\def \g {\gamma}
\def \bi{\bibitem}
\def\a{\alpha }

\def \g {\gamma}

\def \b {\beta}

\def\be{\begin{equation}}
\def\ee{\end{equation}}

\def \bi {\bibitem}
\def \la{\label}

\begin{document}
\date{April 2006}
\begin{titlepage}
\begin{center}

\vspace{2.0cm}
{\Large \bf Killing-Yano equations and $G$-structures}
\\[.2cm]

\vspace{1.5cm}
 {\large   G. Papadopoulos}

 \vspace{0.5cm}
Department of Mathematics\\
King's College London\\
Strand\\
London WC2R 2LS, UK\\

\end{center}

\vskip 1.5 cm
\begin{abstract}

We solve the Killing-Yano equation on manifolds with a $G$-structure for $G=SO(n), U(n), SU(n), Sp(n)\cdot Sp(1), Sp(n), G_2$ and $Spin(7)$.
Solutions include nearly-K\"ahler, weak holonomy $G_2$, balanced
$SU(n)$ and holonomy $G$ manifolds. As an application, we find that particle probes on $AdS_4\times X$ compactifications of type IIA and 11-dimensional
supergravity admit a ${\cal W}$-type of symmetry generated by the fundamental forms.
We also explore the ${\cal W}$-symmetries of string and particle actions in heterotic and  common sector
supersymmetric backgrounds. In the heterotic case, the generators of the ${\cal W}$-symmetries completely characterize the
solutions of the gravitino Killing spinor equation, and the structure constants of the ${\cal W}$-symmetry algebra depend on the solution of the
dilatino Killing spinor equation.

\end{abstract}
\end{titlepage}
\newpage
\setcounter{page}{1}
\renewcommand{\thefootnote}{\arabic{footnote}}
\setcounter{footnote}{0}

\setcounter{section}{0}
\setcounter{subsection}{0}
\newsection{Introduction}

It is known for sometime  that spacetime forms  of various
degrees generate symmetries in particle and string supersymmetric worldvolume  actions.
In string theory, invariance of the worldsheet action requires that these forms are parallel with respect to
a suitable connection leading to special holonomy manifolds \cite{phgp}.
In particular,  the covariant constant forms of Berger  manifolds
are associated with  ${\cal W}$-type of symmetries.
In string theory these ${\cal W}$-type of symmetries are part of the chiral symmetry algebra and
so are essential for the understanding of quantum theory \cite{vafa, deBoer, vid}.

The conditions for the invariance of the action of supersymmetric particle under symmetries generated by spacetime forms are somewhat different.
To describe  these symmetries, let  $X$ a superfield which is a map from
the worldline supermanifold $\Xi^{1|1}$, with coordinates $(t, \theta)$, into the spacetime $M$.  The transformation generated by
a spacetime $(l+1)$-form $\lambda$
is
\bea
\delta X^i=a_l \lambda^i{}_{j_1\dots j_l} DX^{j_1}\dots DX^{j_l}~,
\la{trans}
\eea
where the index is raised using the spacetime metric $g$ and $a_l$ is an infinitesimal parameter.  $D$ is the worldline superspace
derivative $D^2=i\partial_t$.
Requiring that the worldline action\footnote{There are  particle actions with one supersymmetry and additional couplings which can be found in
\cite{coles}.}  of \cite{howe} written in superfields,
\bea
I=-{i\over2} \int dt\, d\theta\, g_{ij}\, DX^i \partial_t X^j~,
\la{act}
\eea
to be invariant under (\ref{trans}), one finds that  the covariant derivative of the form $\lambda$ coincides with
the exterior derivative \cite{holten}, $\nabla\lambda=(l+2)^{-1}d\lambda$, or explicitly,
\bea
\nabla_{i_1} \lambda_{i_2\dots i_{l+2}}=\nabla_{[i_1} \lambda_{i_2\dots i_{l+2}]}~.
\la{invact}
\eea
This condition is  known as Killing-Yano equation \cite{yano}. It has been extensively
investigated in the context of black holes in relation to the integrability of the geodesic motion \cite{penrose, floyd, carter-a},
and in relation to the separation of
Klein-Gordon and Dirac equations \cite{carter-b, chandrasekhar, carter-c}, see also \cite{page}. It is clear that if $\lambda$ is a one-form,
then (\ref{invact}) implies that the associated vector field is Killing.

In this paper, we  initiate the investigation of the question on whether the conditions imposed on the geometry of
a background by the requirement of spacetime supersymmetry can be understood in terms of the conditions
for the existence of symmetries in the worldvolume theories of particle and string probes. This
will give an interpretation of the conditions for spacetime supersymmetry in terms of conditions for existence of worldvolume  Noether symmetries,
and so it may lead to an  understanding of spacetime supersymmetry  in terms of algebras.

It is implicit in the results of \cite{phgp} that the geometry of supersymmetric supergravity backgrounds associated   with
a holonomy $G$ (Berger manifolds),
or equivalently  a parallel $G$-structure,
  can be encoded in terms of
a ${\cal W}$-algebra of a particle or a string worldvolume action. However supersymmetric backgrounds with fluxes exhibit
$G$-structure which are not parallel. {}For this,  we  solve  (\ref{invact}) on  manifolds that
admit a $G$-structure for $G=SO(n)$ $(n)$, $U(n)$ $(2n)$, $SU(n)$ $(2n)$, $Sp(n)\cdot Sp(1)$ $(4n)$,
$Sp(n)$ $(4n)$, $G_2$ $(7)$ and $Spin(7)$ $(8)$, where
in parenthesis is the dimension of the manifold.
This list includes most of the $G$-structures of supersymmetric supergravity backgrounds.
Assuming that $\lambda$ is one of the fundamental forms of these manifolds,
 we shall solve (\ref{invact}).   We shall show
that in most cases (\ref{invact}) implies that  $\lambda$ is parallel with respect to the Levi-Civita connection, ie the manifold is of holonomy $G$.
However,
there are some exceptions. In particular, we find that  nearly K\"ahler, nearly parallel (weak) $G_2$  and balanced $SU(n)$ manifolds arise as solutions.
The results have been tabulated in tables 1, 2 and 3 for the $U(n)$, $SU(n)$ and $G_2$ cases, correspondingly. The remaining are stated at the
appropriate section.

Next, we  examine the relation between the geometric conditions that arise for the existence
of ${\cal W}$-symmetries in world-volume actions and those imposed on a spacetime by the
requirement of spacetime supersymmetry.
   In the heterotic
case, there is a ${\cal W}$-type of symmetry for every form constructed as a bi-linear of the spinors that solve the gravitino Killing spinor equation.
These forms completely characterize the solutions of gravitino Killing spinor equation.
So there is a relation between  the generators of ${\cal W}$ algebra of a probe particle or string
and  the solutions of the gravitino Killing spinor equation. The dilatino Killing spinor equation imposes
conditions on the generators of the ${\cal W}$-algebra that appear in the right-hand-side of commutators of the
symmetries generated by the parallel forms. These are typically vanishing conditions of Nijenhuis type of tensors. Similar observations are made
for common sector backgrounds.
In type II and 11-dimensional supergravity,  nearly K\"ahler and nearly parallel $G_2$ manifolds
appear in $AdS_4\times X$ compactifications, respectively.
One therefore concludes that particle probes on such backgrounds admit ${\cal W}$-symmetries.

This paper is organized as follows: In section two, we solve the Killing-Yano equation on manifolds with a $G$-structure, and
in section three, we explore the applications in the context of supersymmetric supergravity backgrounds.

\newsection{Solving the Killing-Yano equations}

One way to solve  (\ref{invact}) is to use the same technique as that one employs  to classify $G$-structures \cite{gray}.
For all the groups $G$ that are mentioned in the introduction, the Lie algebra of $G$,  $\mathfrak{g}\subseteq \mathfrak{so}(d)=\Lambda^2(\bR^d)$,
where $d={\rm dim}\,M$ is the dimension of the manifold.
One can then write $\Lambda^2(\bR^d)=\mathfrak{g}\oplus\mathfrak{g}^\perp$ and decompose the Levi-Civita connection
 as
\bea
\nabla=\pi(\nabla)+ \sigma(\nabla)
\eea
where $\pi(\nabla)$ takes values in $\mathfrak{g}$ and $\sigma(\nabla)$ in $\mathfrak{g}^\perp$. If $\lambda$ is one of the fundamental
forms of the $G$-structure, and so $G$-invariant,  $\pi(\nabla)\lambda=0$. Thus (\ref{invact}) does not dependent on the $\pi(\nabla)$ components of the connection. So
(\ref{invact}) turns to a condition on $\sigma(\nabla)$.

In practise to carry out these calculations, it is convenient to introduce an adapted frame to the $G$-structure. Then in most cases,
it is straightforward to identify  the components of the frame connection $\Omega$ in $\mathfrak{g}$ and those in $\mathfrak{g}^\perp$, and observe
that (\ref{invact}) depends only on the latter. The conditions that (\ref{invact}) imposes on $\sigma(\Omega)$ can then be investigated in each case.

If $G=SO(n)$, $d=n$, the fundamental form is the  volume form of the Riemannian metric. This form is parallel
with respect to the Levi-Civita connection and so (\ref{invact}) is satisfied without further restrictions.

\subsection{$U(n)$}

Let  $M$, $ d=2n$, be a manifold with  structure group $U(n)$. The fundamental form
is the Hermitian form $\omega(X,Y)= g(X, IY)$  of  an almost complex structure $I$ with compatible Hermitian metric $g$.
In this case $\lambda$ can be identified  with $\omega$ or its skew-symmetric powers $\omega^k=\wedge^k\omega$.

To identify $\sigma(\Omega)$ in this case choose a Hermitian frame, ie $g=\delta_{ij} e^i e^j=2\delta_{\a\bar\b} e^\a e^{\bar \b}$ and
$\omega=-i \delta_{\a\bar\b} e^\a e^{\bar \b}$. Then the $\sigma(\Omega)$ components of the frame connection $\Omega$
are $\Omega_{i,\a\b}$ and their complex conjugates. Observe that (\ref{invact}) does not depend on the
components $\Omega_{i,\a\bar\b}$ which are along $\mathfrak{u}(n)$. The different $U(n)$ structures
can be found by decomposing $\sigma$ into four irreducible representations $W_1,\dots W_4$ - the Gray-Hervella classes \cite{gray}.
After some computation one finds that  the invariance condition (\ref{invact}) implies that  all
 these classes vanish apart from $W_1$, ie $W_2=W_3=W_4=0$.
The Nijenhuis tensor ${\cal N}(I)$ of such manifolds with a $U(n)$ structure does not vanish.
In particular, it is a (3,0) and (0,3),
 and $\nabla$-parallel form. $M$ is a nearly K\"ahler manifold. If ${\rm dim}\, M=4$, then  $M$ is K\"ahler.
 If ${\rm dim}\, M= 6$ and M is not  K\"ahler,  then the structure group reduces to $SU(3)$. An example of such a
 manifold is $S^6$. Similarly for ${\rm dim}\, M\geq 6$ and $M$  not K\"ahler,
 the structure group reduces to a
 proper subgroup of $U(n)$.

Next if $\lambda=\omega^k$,  $1<k<n$ a rather lengthy but straightforward computation reveals that (\ref{invact}) implies that
$M$ is K\"ahler. The conditions on the
geometry of $M$ are summarized in the table 1.

\begin{table}[ht]
 \begin{center}
\begin{tabular}{|c|c|c|}\hline
 ${\rm Symmetries}$ & {\rm Conditions}&{\rm Geometry}
 \\ \hline \hline
 $\omega$& $W_2=W_3=W_4=0$ & {\rm Nearly K\"ahler}
\\ \hline
 $\omega^k~~~1<k<n$&  $W_1=W_2=W_3=W_4=0$& {\rm  K\"ahler}\\
\hline
\end{tabular}
\end{center}
\caption{The  columns give the forms that are associated with the symmetries and the conditions for the
transformations to leave the action invariant, respectively.}
\end{table}

\subsection{$SU(n)$}
The fundamental $SU(n)$ forms are the Hermitian form $\omega$, as in the $U(n)$ case, and the real, ${\rm Re}\chi$, and
imaginary,  ${\rm Im}\chi$,  parts
of a (n,0)-form $\chi$. Adapting a Hermitian frame to the $SU(n)$ structure, $\sigma$ is spanned by the
component $\Omega_{i, \a}{}^\a$ of the frame connection in addition to those described for the $U(n)$ case.
In fact $\sigma$ is now decomposed in five $SU(n)$ irreducible representations $W_1,\dots, W_5$ - the $SU(n)$ classes \cite{salamon}.

 The investigation of the symmetries generated  with the Hermitian form $\omega$ and its skew-symmetric powers
is identical to that we have done for the $U(n)$ case. The conditions on the geometry are as those of the $U(n)$ case above.

In the $SU(n)$ case, additional transformations can be constructed from the  real ${\rm Re}\chi$ and imaginary  ${\rm Im}\chi$
components of the $(n,0)$-form $\chi$. For these transformations to be symmetries of the action (\ref{act}), the invariance condition (\ref{invact})
 implies
that $W_4=W_5=0$ provided that $n>2$. The $SU(n)$ classes have been chosen as in \cite{gpug}.
For $n=1$, the analysis is identical to the $Sp(1)$ case that we shall explain below.
The results are summarized in the table 2.

\begin{table}[ht]
 \begin{center}
\begin{tabular}{|c|c|c|}\hline
 ${\rm Symmetries}$ & {\rm Conditions}&{\rm Geometry}
 \\ \hline \hline
 $\omega$& $W_2=W_3=W_4=0$ & {\rm Nearly K\"ahler}
\\ \hline
 $\omega^k~~~1<k<n$&  $W_1=W_2=W_3=W_4=0$& {\rm  K\"ahler}\\
\hline
 ${\rm Re}\chi\,,~~{\rm Im}\chi$&  $W_4=W_5=0$& {\rm Balanced Hermitian}\\
\hline
$\omega\,,~~{\rm Re}\chi\,,~~{\rm Im}\chi$&  $W_2=W_3=W_4=W_5=0$& {\rm Special nearly K\"ahler}\\
\hline
$\omega^k~~~1<k<n\,,~~{\rm Re}\chi\,,~~{\rm Im}\chi$&  $W_1=W_2=W_3=W_4=W_5=0$& {\rm  Calabi-Yau}\\
\hline

\end{tabular}
\end{center}
\caption{The  columns give the forms that are associated with the symmetries and the conditions for the
transformations to leave the action invariant, respectively.}
\end{table}

\subsection{$Sp(n)\cdot Sp(1)$}

The $Sp(n)\cdot Sp(1)$  fundamental is
\bea
\chi=\wedge^2\omega_I+\wedge^2\omega_J+\wedge^2 \omega_K~,
\eea
where $\omega_I, \omega_J$ and $\omega_K$ are locally defined  Hermitian forms associated with
a quaternionic structure $I,J$ and $K$, $K=IJ$.

To solve (\ref{invact}) for $\lambda=\chi$, we adapt a Hermitian frame with respect to the $I$
almost complex structure. Then $\omega_J$ becomes a locally defined (2,0) and (0,2) form.
The components of  frame connection along $\sigma$ are spanned by
$\mathring{\Omega}_{i,\a\b}$ and $\mathring{A\Omega}_{i,\a\bar\beta}$, where
$\mathring{\Omega}_{i,\a\b}$ is the $\omega_J$ traceless part of $\Omega_{i,\a\b}$ and $\mathring{A\Omega}_{i,\a\bar\beta}$
is the component of $\Omega_{i,\a\bar\b}$ which satisfies
\bea
\mathring{A\Omega}_{i,\a\bar\beta}\, J^{\bar\beta}{}_\gamma=-\mathring{A\Omega}_{i,\gamma\bar\beta}\, J^{\bar\beta}{}_\a~,~~~
\mathring{A\Omega}_{i,\a}{}^\a=0~.
\eea
The components of $\nabla$ along the $\mathfrak{sp}(n)\oplus \mathfrak{sp}(1)$ directions are spanned by the remaining components
of the frame connection. In particular the $\mathfrak{sp}(1)$ directions are spanned by the $\omega_J$ trace of $\Omega_{i,\a\b}$
and the $\omega_I$ trace of $\Omega_{i,\a\bar\b}$. It is known that if $n>2$,  $\sigma$ decomposes in six $Sp(n)\cdot Sp(1)$ irreducible representations,
and that if $n=2$,  $\sigma$ decomposes in four irreducible representations \cite{swann}. It is also known that all these classes are determined
by evaluating $\nabla\chi$.

The invariance condition (\ref{invact}) imposes conditions on $\sigma$, ie on  $\mathring{\Omega}_{i,\a\b}$ and on
$\mathring{A\Omega}_{i,\a\bar\beta}$.  After a long but straightforward computation, one finds that (\ref{invact}) implies that $\sigma$ vanishes.
Thus
$\chi$ is parallel and invariance of the action implies that $M$  is quaternionic K\"ahler manifold.

\subsection{$Sp(n)$}

The $Sp(n)$ fundamental forms are the  Hermitian forms  $\omega_I$, $\omega_J$ and $\omega_K$ of an almost
hypercomplex structure $I,J$ and $K$, $K=IJ$.
The conditions for the transformations  generated by $\wedge^k\omega_I$, $\wedge^k\omega_J$ and $\wedge^k\omega_k$, separately, to be symmetries
are exactly as those
we have found for the $U(n)$ case associated with $I$ or $J$ or $K$ almost complex structures.

We shall also investigate the conditions that are implied by taking $\lambda$ to be $\omega_I$ and $\omega_J$.
For this we introduce a Hermitian frame with respect to $I$ as in the $Sp(n)\cdot Sp(1)$ case above. However
in this case, the components of the frame connection along $\sigma$ are spanned by $\mathring{\Omega}_{i,\a\b}$ and $\mathring{A\Omega}_{i,\a\bar\beta}$
and the components along the $\mathfrak{sp}(1)$ directions, ie they are spanned by $\Omega_{i,\a\b}$ and ${A\Omega}_{i,\a\bar\beta}$.
For $\omega_I$ to satisfy (\ref{invact}), the non-vanishing components of the frame connection are $\Omega_{[\a,\b\g]}$.
It remains to find the additional  conditions that (\ref{invact}) imposes for $\lambda=\omega_J$. A straightforward computation reveals that
$\Omega_{[\a,\b\g]}=0$. So (\ref{invact}) for $\lambda=\omega_I$ and  $\lambda=\omega_J$ implies that $M$ is a hyper-K\"ahler manifold.

\subsection{$G_2$}

The transformations are generated  by  either the $G_2$ fundamental 3-form  $\varphi$ or its dual $\star\varphi$, or both.
It is known that  $\Lambda^2(\bR^7)$ decomposes in $G_2$ representations as $\Lambda^2(\bR^7)=\mathfrak{g}_2 \oplus {\bf 7}$. So the $\sigma$
 directions
of the Levi-Civita connection are along the 7-dimensional representation.
Adapting an appropriate frame, the $\sigma$ directions of the frame connection can be written
as
\bea
\sigma(\Omega)_{i,jk}= L_{im} \varphi^m{}_{jk}~.
\la{scop}
\eea
It is known that the $G_2$ structures on a 7-dimensional manifold can be characterized by four classes \cite{fgray}. Schematically, one
has
\bea
\nabla\varphi\Longleftrightarrow X_1+X_2+X_3+X_4
\eea
where $X_1$ is a singlet, $X_2$ is in ${\bf 14}$ representation, $X_3$ is in the ${\bf 27}$ representation and $X_4$ is in the
${\bf 7}$ representation.

As we have explained (\ref{invact}) depends only on the $\sigma$ component (\ref{scop}) of the frame connection.
After some computation, one can show that $\varphi$ solves (\ref{invact}),  iff $L_{ij}=f \delta_{ij}$. This in turn implies that
$X_2=X_3=X_4=0$, ie $M$ is nearly parallel or weak $G_2$ manifold. For such $G_2$
manifolds
\bea
d\varphi=X_1 {}\,\,\star\varphi~.
\eea
A similar computation reveals that $\star\varphi$ generates a symmetry iff $M$ is holonomy $G_2$. The results have been tabulated in
table 3.
\begin{table}[ht]
 \begin{center}
\begin{tabular}{|c|c|c|}\hline
 ${\rm Symmetries}$ & {\rm Conditions}&{\rm Geometry}
 \\ \hline \hline
 $\varphi$& $X_2=X_3=X_4=0$&{\rm Nearly Parallel}
\\ \hline
 $\star\varphi$&  $X_1=X_2=X_3=X_4=0$&{\rm Holonomy}\,\, $G_2$\\
\hline
$\varphi,~ \star\varphi$& $X_1=X_2=X_3=X_4=0$&{\rm Holonomy}\,\, $G_2$
\\ \hline
\end{tabular}
\end{center}
\caption{The  columns give the forms that are associated with the symmetries and the conditions for the
transformations to leave the action invariant, respectively.}
\end{table}

\subsection {$Spin(7)$}

The transformation in this case is generated by the fundamental $Spin(7)$ self-dual 4-form $\phi$. Observe that
(\ref{invact}) implies that $\phi$ is co-closed. Since $\phi$ is self-dual is also closed. In turn, (\ref{invact}) implies that it is also parallel.
 Therefore
$\phi$ generates a symmetry iff $M$ is a holonomy $Spin(7)$ manifold.

\newsection{Supersymmetric backgrounds and {\cal W}-symmetries}

\subsection{Heterotic and common sector backgrounds}

The geometry of all supersymmetric heterotic string backgrounds has been described in \cite{het}.
Such backgrounds admit $\hat\nabla$-parallel forms of various degrees. These are constructed as form bilinears
of the spinors $\hat\epsilon$ that solve the gravitino Killing spinor equation $\hat\nabla\hat\epsilon=0$,   where $\hat\nabla$ is a metric connection
with skew-symmetric torsion\footnote{For conventions and
notation see \cite{gpug}.}. These forms have been given in \cite{het}.

The relevant part of heterotic string worldsheet Lagrangian is
\bea
L=(g+b)_{ij} D_+X^i \partial_= X^j
\la{acta}
\eea
where $X$ is a (1,0) superfield. This Lagrangian  apart from the metric coupling it also contains a Wess-Zumino term.
The transformation generated by a spacetime form $\lambda$ is
\bea
\delta X^i=a_l \lambda^i{}_{j_1\dots j_l} D_+X^{j_1}\dots D_+X^{j_l}~.
\la{strans}
\eea
This transformation leaves the action invariant, iff  $\lambda$ is $\hat\nabla$-parallel, $\hat\nabla\lambda=0$ \cite{phgp}.
Thus all forms constructed as $\hat\nabla$-parallel spinor bi-linears generate worldsheet ${\cal W}$-symmetries.
So in this case there is a direct correspondence between the conditions on the spacetime  that arise from the solution of
 the gravitino Killing spinor
equation and those that arise  for the existence of ${\cal W}$-symmetries in string world-sheet actions.

The commutator of two ${\cal W}$-symmetries (\ref{trans}) generated by $\lambda$ and $\mu$
has been given in \cite{phgp}. Typically, the ${\cal W}$-algebra does not close
on the original transformations of the form (\ref{strans}),  and new generators can arise. Some of these new generators
are transformations of the type (\ref{strans}) associated with the Nijenhuis tensor ${\cal N}(\lambda, \mu)$.
The dilatino Killing spinor equation of the heterotic string imposes restrictions on these Nijenhuis tensors \cite{het}.
In particular in many cases, they are required to vanish.
As a result, the conditions imposed by the dilatino Killing spinor equations can be interpreted as restrictions
on the structure constants of the ${\cal W}$-algebra.

A similar analysis can be made for the supersymmetric backgrounds of the common sector, see also \cite{het, phgp}. The geometry of
supersymmetric
common sector backgrounds is not as well understood as that of the heterotic ones. For example the gravitino Killing spinor
equation\footnote{We follow the notation of \cite{gpug}.}, which
is two copies of the heterotic one, $\hat\nabla\hat\epsilon=0$ and $\check\nabla\check\epsilon=0$,
has not been solved in all cases for more than two supersymmetries \cite{gpug}.
The analysis for the form bilinears constructed for either the $\hat\epsilon$  or the
$\check\epsilon$  parallel spinors is the same as that already given for the heterotic string.
Supersymmetric common sector backgrounds admit additional form bilinears constructed from a $\hat\epsilon$ and a $\check\epsilon$
parallel spinor. Such bilinears are not parallel with respect to either $\hat\nabla$ or $\check\nabla$ connections.
Requiring that these bilinears generate worldsheet ${\cal W}$-symmetries will impose additional conditions
on the geometry of spacetime to those associated with gravitino Killing spinor equation.

The relevant worldline action suitable to investigating ${\cal W}$-symmetries for a particle in a
heterotic or common sector background is that of \cite{coles} which contains an additional skew-symmetric coupling for the
fermions. One can show that forms that are parallel with respect to a connection with skew-symmetric torsion
generate ${\cal W}$-symmetries for the worldline action. The analysis is similar to that for the heterotic string and the common sector
above.

\subsection{Type II backgrounds}

The particle and string actions that probe generic type II backgrounds are of Green-Schwarz type. So the
fermions are spacetime spinors rather than worldvolume spinors which transform as spacetime vectors, ie as those in (\ref{act}) and (\ref{acta})
which we have used to construct the ${\cal W}$-symmetries.
Green-Schwarz actions can be rewritten in terms of worldvolume fermions after a gauge fixing procedure to eliminate additional
degrees of freedom, see eg \cite{gsw}. However, this done on a case by case basis and there is not a general procedure for generic supersymmetric
type II backgrounds. So we cannot directly compare the geometry of supersymmetric type II backgrounds and the conditions
that arise from the invariance of worldvolume actions under transformations generated by spacetime forms.

Nevertheless supersymmetric type II and 11-dimensional supergravity backgrounds admit forms of various
degrees constructed from Killing spinor bi-linears, see e.g. \cite{pakis, ggp, ggpb}. From the results of section 2 whenever the backgrounds
are of holonomy $G$,  the ${\cal W}$-symmetries have been described in \cite{phgp}. For the remaining cases,
it is known that some $AdS_4\times X$ compactifications of 11-dimensional supergravity require that $X$ to be a nearly parallel
$G_2$ manifold, \cite{romans}, see also \cite{josehull}. This condition coincides with the condition given in table 3
for the existence of a ${\cal W}$-symmetry.
Similarly, $AdS_4\times X$ compactifications of IIA supergravity require that $X$ is a nearly K\"ahler manifold
which coincides with the condition for the existence of a ${\cal W}$-symmetry in tables 1 and 2. Therefore,
we conclude that particle probes of such compactification exhibit ${\cal W}$-symmetries.

In general, the conditions that one obtains from solving the Killing spinor equations of type II and 11-dimensional
backgrounds are, so far, less stringent from those obtained by requiring invariance of worldvolume actions under a transformation
generated by a spacetime form. This may change provided that the ${\cal W}$-symmetries are formulated directly for
Green-Schwarz actions. Consideration should also be given to probes that their dynamics is described
in terms of field equations. In such a case, it has been shown that the requirement of worldvolume supersymmetry puts less restriction
on the spacetime geometry
\cite{stelle}. For the particle case, the action (\ref{act}) that we have used for the analysis can be extended
to include an additional 3-form and other couplings \cite{coles}. Then the conditions for the existence of ${\cal W}$-symmetries
change, see \cite{sfetsos,gibstelpap}. In particular, they lead to a modification of   Killing-Yano equations. The restrictions that these modified
 conditions
impose on the geometry and their relation to those found by the requirement of spacetime supersymmetry on a spacetime will be investigated elsewhere.

\setcounter{section}{0}

\end{document}